\documentclass{aa}
\usepackage{txfonts,psfig,graphicx,natbib,aalongtable}

\begin{document}
   \title{Visible spectroscopy of the new ESO large programme on trans-Neptunian objects and Centaurs: final results\thanks{Based on observations obtained at the VLT Observatory Cerro Paranal of European Southern Observatory, ESO, Chile, in the framework of programs 178.C-0036.}}
 \author{S. Fornasier           \inst{1,2}\and
         M.A. Barucci            \inst{1}\and
         C. de Bergh           \inst{1}\and
         A. Alvarez-Candal            \inst{1,3}\and
         F. DeMeo                 \inst{1}\and
         F. Merlin                \inst{1,4}\and
         D. Perna                 \inst{1,5,6}\and 
         A. Guilbert              \inst{1}\and
         A. Delsanti               \inst{1}\and
         E. Dotto                  \inst{6}\and
         A. Doressoundiram        \inst{1}
        }
   \offprints{}

   \institute{LESIA, Observatoire de Paris, 5 Place J. Janssen, 92195 Meudon Pricipal Cedex, France \\
              \email{sonia.fornasier@obspm.fr}
         \and Universit\'{e} de Paris 7 {\it Denis Diderot},  4, rue Elsa Morante, F-75013 Paris, France
    \and European Southern Observatory, Casilla 19001, Santiago, Chile 
    \and Department of Astronomy, University of Maryland, College Park, MD 20742, USA 
    \and Universit\`a di Roma Tor Vergata, Italy 
    \and INAF-Osservatorio Astronomico di Roma, via Frascati 33, 00040 Monteporzio Catone (Roma), Italy
}

   \date{Received May 2009}

\abstract{A second large programme (LP) for the physical studies of TNOs and Centaurs, started at ESO Cerro Paranal
   on October 2006 to obtain high-quality data, has recently been concluded.  In this paper
we present the spectra of these pristine bodies obtained in the visible range during the last two semesters (November 2007--November 2008) of the LP.}{ We investigate the spectral behaviour of the TNOs and Centaurs observed, and we analyse the spectral slopes distribution of the full data set coming from this LP and from the literature.}
{Spectroscopic observations in the visible range were carried out at the UT1 (Antu) telescope using the instrument FORS2. We computed the spectral slope for each observed object, and searched for possible weak absorption features. A statistical analysis was performed on a total sample of 73 TNOs and Centaurs to look for possible correlations between dynamical classes, orbital parameters, and spectral gradient. }
{We obtained new spectra for 28 bodies (10 Centaurs, 6 classical, 5 resonant, 5 scattered disk, and 2 detached objects), 15 of which were observed for the first time. All the new presented spectra are featureless, including 2003 AZ84, for which a faint and broad absorption band possibly attributed to hydrated silicates on its surface has been reported (Fornasier et al. 2004a, and Alvarez-Candal et al. 2008). The data confirm a wide variety of spectral behaviours, with neutral--grey to very red gradients. An analysis of the spectral slopes available from this LP and in the literature for a total sample of 73 Centaurs and TNOs shows that there is a lack of very red objects in the classical population. 
We present the results of the statistical analysis of the spectral slope distribution versus orbital parameters. In particular, we confirm a strong anticorrelation between spectral slope and orbital inclination for the classical population. Nevertheless, we do not observe a change in the slope distribution at $i \sim 5^{o}$, the boundary between the dynamically hot and cold populations, but we find that objects with $ i < 12^{o}$ show no correlation between spectral slope and inclination, as already noticed by Peixinho et al. (2008) on the colour --inclination relation for classical TNOs. A strong correlation is also found between the spectral slope and orbital eccentricity for resonant TNOs, with objects having higher spectral slope values with increasing eccentricity.}{}

   \keywords{Centaurs -- TNOs  -- visible -- spectroscopy }

\titlerunning{Visible spectroscopy of TNOs and Centaurs}
\authorrunning{Fornasier et al.}
   \maketitle
%

\section{Introduction}
The icy bodies in orbit beyond Neptune (trans-Neptunians or TNOs) and Centaurs represent the most primitive population of all the Solar System objects. They are fossil remnants of the formation of our planetary system and the investigation of their surface properties is essential for understanding the formation and the evolution of the Solar System.
Since 1992, more than 1300 TNOs and Centaurs have been detected.
The TNO population is dynamically classified into several categories (Gladman et al. 2008): classical objects have orbits with low eccentricities and 
semi-major axes between about 42 and 48 AU; resonant objects are trapped in resonances with Neptune, the majority of them located in or near the 3:2 mean motion
resonance; scattered objects have high-eccentricity, high-inclination orbits and a perihelion distance near q=35 AU; extended scattered disk (or detached) objects (SDOs) are located at distances so great that they cannot have been emplaced by gravitational
interactions with Neptune. In addition, the Centaurs are closest to the Sun and have unstable orbits between those of Jupiter and Neptune. They seem to originate from the Kuiper Belt and should have been injected into their present orbits by perturbations from giant planets or mutual collisions.

The investigation of the surface composition of these icy bodies can provide essential information about the conditions in the early Solar System environment at large distances from the Sun. \\
Aiming at obtaining high--quality data of these populations, a large programme (LP) for the
observation of Centaurs and TNOs was started using the facilities of the ESO / Very Large
Telescope site at Cerro Paranal in Chile (PI: M.A. Barucci). More than 500 hours were allocated to observe these objects in visible and near-infrared photometry--spectroscopy, and V--R polarimetry from November 2006 until December 2008.

In this paper we report the results of the visible spectroscopy of TNOs and Centaurs performed with the instrument FORS2 at the VLT unit 1 {\it Antu} during the last two semesters of the LP (November 2007-November 2008). New visible spectra of 28 objects with $ 18.7 < V < 21.8$, were acquired, 15 of which were observed for the first time. A statistical analysis of the full data set of spectral slopes coming from this LP and from the literature is presented on a total sample of 73 TNOs and Centaurs.

\begin{table*}[t]
\caption{Observational conditions of the TNOs and Centaurs spectroscopically investigated. }
{\footnotesize
\label{tab_obs}
\begin{center}
\begin{tabular}{|l|l|l|c|c|c|c|c|l|c|} \hline
{\bf  Object} & {\bf Date} & {\bf UT} &  {\boldmath{$\Delta$}} & {{\boldmath $r$}} & {\boldmath $\alpha$} &{\bf T\boldmath{$_{exp}$}} & {\bf airm.} & {\bf Solar An. (airm.)} & {\bf Slope} \\ 
& & &   {\bf [AU]} & {\bf [AU]} &  {\boldmath [$^{o}$]} & {\bf [s]} &  & & {\bf [\%/(\boldmath{$10^{3}$ \AA)}]} \\ \hline
{\em CENTAURS} & & & & & & & & & \\ 
5145 Pholus        & 2008 Apr. 12  & 05:57:44 & 21.214 & 21.864 & 2.0 & 2600 & 1.21 & Ld102$-$1081 (1.15) &  48.6$\pm$0.7 \\
10199 Chariklo     & 2008 Feb. 04  & 07:57:47 & 13.301 & 13.395 & 4.2 & 900 & 1.10 & Ld102$-$1081 (1.13)  &  10.4$\pm$0.6 \\
52872 Okyrhoe      & 2008 Feb. 04  & 06:33:30 &  4.879 &  5.800 & 3.8 & 1800 & 1.14 & Ld102$-$1081 (1.13)  &  13.4$\pm$0.6\\
55576 Amycus       & 2008 Apr. 12  & 03:43:14 & 15.206 & 16.056 & 2.0 & 2400 & 1.16 & Ld102$-$1081 (1.15)  &  37.1$\pm$0.9 \\
{\bf 120061 (2003 CO1)}  & 2008 Apr. 12  & 04:48:44 & 10.180 & 11.122 & 1.8 & 2400 & 1.13 & Ld102$-$1081 (1.15)  &  9.4$\pm$0.6  \\
{\bf 2002 KY14}          & 2008 Sept. 21 & 01:11:40 &  7.802 &  8.649 & 3.8 & 1200 & 1.25 & Ld112$-$1233(1.12)   & 36.9$\pm$0.7  \\
{\bf 2007 UM126}         & 2008 Sept. 21 & 04:46:01 & 10.202 & 11.163 & 1.6 & 2400 & 1.10 & Ld112$-$1233(1.12)   &  8.8$\pm$0.7 \\ 
{\bf 2007 VH305}         & 2008 Nov. 22  & 02:38:07 &  7.853 &  8.638 & 4.2 & 2400 & 1.18 & Ld98$-$978 (1.20)   & 12.1$\pm$0.7 \\
{\bf 2008 FC76}          & 2008 Sept. 21 & 00:21:29 & 10.976 & 11.688 & 3.6 & 1800 & 1.29 & Ld112$-$1233(1.12)   & 36.0$\pm$0.7 \\
{\bf 2008 SJ236}         & 2008 Nov. 22  & 00:53:58 &  5.522 &  6.364 & 5.0 & 4000 & 1.28 & Ld93$-$101 (1.30)   & 20.9$\pm$0.8 \\ \hline
{\em CLASSICALS} & & & & & & & & & \\ 
{\bf 20000 Varuna}       & 2007 Dec. 04  & 06:41:41 & 42.606 & 43.392 & 0.8 & 2400 & 1.57 & Hyades64 (1.41)      &  24.9$\pm$0.6     \\
120178 (2003 OP32) & 2008 Sept. 21 & 02:26:48 & 40.545 & 41.365 & 0.8 & 2400 & 1.14 & Ld112$-$1233(1.12)   &  0.9$\pm$0.7  \\ 
{\bf 120348 (2004 TY364) }& 2008 Nov. 22  & 04:53:01 & 38.839 & 39.591 & 0.9 & 2400 & 1.27 & Ld93$-$101 (1.30)   & 22.9$\pm$0.7 \\
{\bf 144897 (2004 UX10)} & 2007 Dec. 04  & 01:10:32 & 38.144 & 38.836 & 1.0 & 2400 & 1.13 & HD1368 (1.12)        &  20.7$\pm$0.8  \\
{\bf 144897 (2004 UX10)} & 2007 Dec. 05  & 02:41:59 & 38.158 & 38.836 & 1.1 & 2400 & 1.19 & Ld93101 (1.24)       &  19.2$\pm$0.9  \\
145453 (2005 RR43) & 2007 Dec. 04  & 02:59:10 & 37.623 & 38.511 & 0.6 & 2400 & 1.12 & HD1368 (1.12)        &   1.6$\pm$0.6   \\ 
{\bf 174567 (2003 MW12)} & 2008 Apr. 12  & 07:17:41 & 47.318 & 47.968 & 0.9 & 3700 & 1.09 & Ld102$-$1081 (1.10)  &  19.2$\pm$0.6 \\ \hline
{\em RESONANTS} & & & & & & & & & \\
{\bf 42301 (2001 UR163)}  & 2007 Dec. 05  & 01:10:23 & 49.625 & 50.308 & 0.8 & 2400 & 1.23 & Ld93$-$101 (1.24)    &  50.9$\pm$0.7     \\
55638 (2002 VE95)  & 2007 Dec. 05  & 04:13:26 & 27.297 & 28.248 & 0.5 & 2400 & 1.21 & Ld93$-$101 (1.24)    &  40.0$\pm$0.7      \\
90482 Orcus        & 2008 Feb. 04  & 05:25:24 & 46.901 & 47.807 & 0.5 & 2400 & 1.07 & Ld102$-$1081 (1.11)  &   1.6$\pm$0.6  \\
208996 (2003 AZ84) & 2008 Nov. 22  & 06:29:23 & 44.883 & 45.458 & 1.0 & 2800 & 1.30 & Ld93$-$101 (1.30)   & 3.6$\pm$0.6 \\ 
{\bf 2003 UZ413}         & 2007 Dec. 04  & 04:42:24 & 41.171 & 42.004 & 0.7 & 2400 & 1.36 & Hyades64 (1.41)      &   6.2$\pm$0.6  \\
{\bf 2003 UZ413}         & 2008 Sept. 21 & 05:46:16 & 41.469 & 42.163 & 1.0 & 2400 & 1.25 & Ld112$-$1233(1.12)   &  5.0$\pm$0.8  \\ 
{\bf 2003 UZ413}         & 2008 Nov. 22  & 03:49:31 & 41.275 & 42.197 & 0.5 & 2400 & 1.15 & Ld98$-$978 (1.17)   & 4.9$\pm$0.7 \\ \hline
{\em SDOs} & & & & & & & & & \\ 
42355 Typhon       & 2008 Apr. 12  & 01:07:02 & 16.892 & 17.650 & 2.2 & 1200 & 1.22 & Ld102$-$1081 (1.15)  &  12.1$\pm$0.8 \\
44594 (1999 OX3)   & 2008 Sept. 21 & 03:22:39 & 22.023 & 22.889 & 1.3 & 2400 & 1.07 & Ld110$-$361 (1.10)   & 36.2$\pm$0.8  \\ 
{\bf 73480  (2002 PN34)} & 2007 Nov. 10  & 00:51:51 & 14.893 & 15.344 & 3.3 & 2800 & 1.22 & LD115$-$871 (1.13)   &  15.8$\pm$0.7 \\
{\bf 145451 (2005 RM43)} & 2007 Dec. 04  & 04:04:16 & 34.298 & 35.195 & 0.7 & 2400 & 1.15 & HD1368 (1.12)        &   2.2$\pm$0.7    \\
{\bf 2007 UK126}         & 2008 Sept. 21 & 08:18:31 & 45.136 & 45.619 & 1.1 & 2400 & 1.07 & Ld112$-$1233(1.12)   & 19.6$\pm$0.7 \\ \hline
{\em DETACHED OBJECTS} & & & & & & & & & \\ 
90377 Sedna        & 2008 Sept. 21 & 06:53:28 & 87.419 & 88.015 & 0.5 & 3600 & 1.20 & Ld112$-$1233(1.12)   & 40.2$\pm$0.9 \\
120132 (2003 FY128)& 2008 Apr. 12  & 02:32:57 & 37.477 & 38.454 & 0.3 & 2800 & 1.06 & Ld102$-$1081 (1.10)  &  26.7$\pm$1.0  \\ \hline
\end{tabular}
\end{center}
}
\begin{list}{}{}
\item Observational conditions: for each object we report the observational date and universal time (UT of the beginning of the
exposure), the geocentric distance ($\Delta$), the heliocentric distance ($r$), the phase angle ($\alpha$), 
the total exposure time, the airmass (mean of the airmass value at the beginning and
at the end of observation), the observed solar analogue stars with their airmass used
to remove the solar contribution, and the spectral slope value computed in the 0.5--0.8 $\mu$m range. The TNOs and Centaurs observed in spectroscopy for the first time are in bold.
\end{list}
\end{table*}

\section{Observations and data reduction}

The data were obtained primarily during five runs in visitor mode on December 2007 and February, April, September, and November 2008. Only one object, 73480 (2002 PN34), was observed in service mode on November 10, 2007. 
 All the details on the spectroscopic observations are given in Table~\ref{tab_obs}.
The spectra were obtained using a low-resolution grism (150 grooves/mm) with a
1 arcsec wide slit, covering the 4400--9300 \AA\ wavelength range with a spectral resolution of about 200.
The slit was oriented along the parallactic angle to minimise the effects of atmospheric differential refraction, which is also corrected by a system of two silica prisms up to an airmass of 1.4. The FORS2 detector is a mosaic of two 2K$\times$4K MIT/LL CCDs with pixel size 15 $\mu$m, corresponding to a pixel scale of 0.126''/px), used in a 2$\times$2 binned mode and with the high-gain read-out mode (1.45 e$^-$/ADU). 

During each night we also acquired 
bias, flat--field, calibration lamp (He-Ar), and several solar analogue star spectra at different
intervals throughout the night. Spectra were reduced using normal data reduction procedures (see Fornasier et al., 2004a) with
the software package Midas.  The wavelength calibration was performed using helium, HgCd, and argon lamp spectral lines. The reflectivity of the objects was obtained by dividing their spectra by the
spectrum of the solar analogue star closest in time and airmass, as reported in Table~\ref{tab_obs}. \\
Spectra have been normalised to unity at 5500 \AA\ and finally smoothed with a median filter technique, using a
box width of 39 \AA\ in the spectral direction for each point of the spectrum. Threshold was set to 0.1,
meaning that the original value was replaced by the median value when this last differed by more than 
10\% from the original one. Finally, for each object we computed the slope $S$ of the spectral continuum
using a standard least squared technique for a linear fit in the
wavelength range between 5000 and 8000 \AA.
The computed slopes and errors are listed in Table~\ref{tab_obs}. 
The reported error bars take the
$1\sigma$ uncertainty of the linear fit into account plus 0.5\%/($10^3$\AA)
attributable to the use of different solar analogue stars.

Each night, immediately before spectral measurements, photometric data in the $V-R-I$ filters (except during the November and December 2007 runs, in which I filter observations were not taken) were obtained for each target, together with almost simultaneous photometric and spectroscopic observations in the near infrared with ISAAC and SINFONI instruments at VLT units 1 {\it Antu} and 4 {\it Yepun}. The results on V+NIR photometry are presented in Perna et al. (2009a), while the near infrared spectroscopic data are still under analysis. \\

The new spectra of the Centaurs and TNOs are shown in Fig.~\ref{f1},~\ref{f2},~\ref{f3}, ~\ref{f4}, and ~\ref{f5}. In these figures we also plot the $V-R$ and $V-I$ (when available) colour indices converted to relative reflectance, colours that show nice agreement with the spectral gradient.

\section{Results}

\addtocounter{table}{1}

\begin{figure}
   \centering
\includegraphics[width=7cm,angle=90]{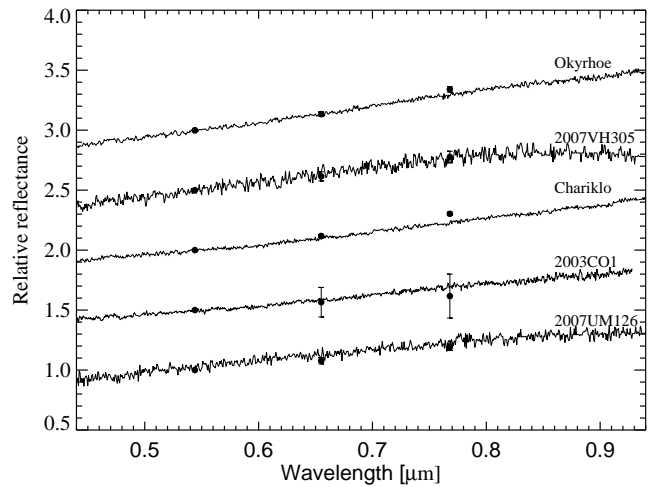}
      \caption{Visible spectra of Centaurs. Spectra are shifted by 0.5 for clarity. The colour indices converted to spectral reflectance are also shown on each spectrum.}
         \label{f1}
   \end{figure}

\begin{figure}
   \centering
\includegraphics[width=7cm,angle=90]{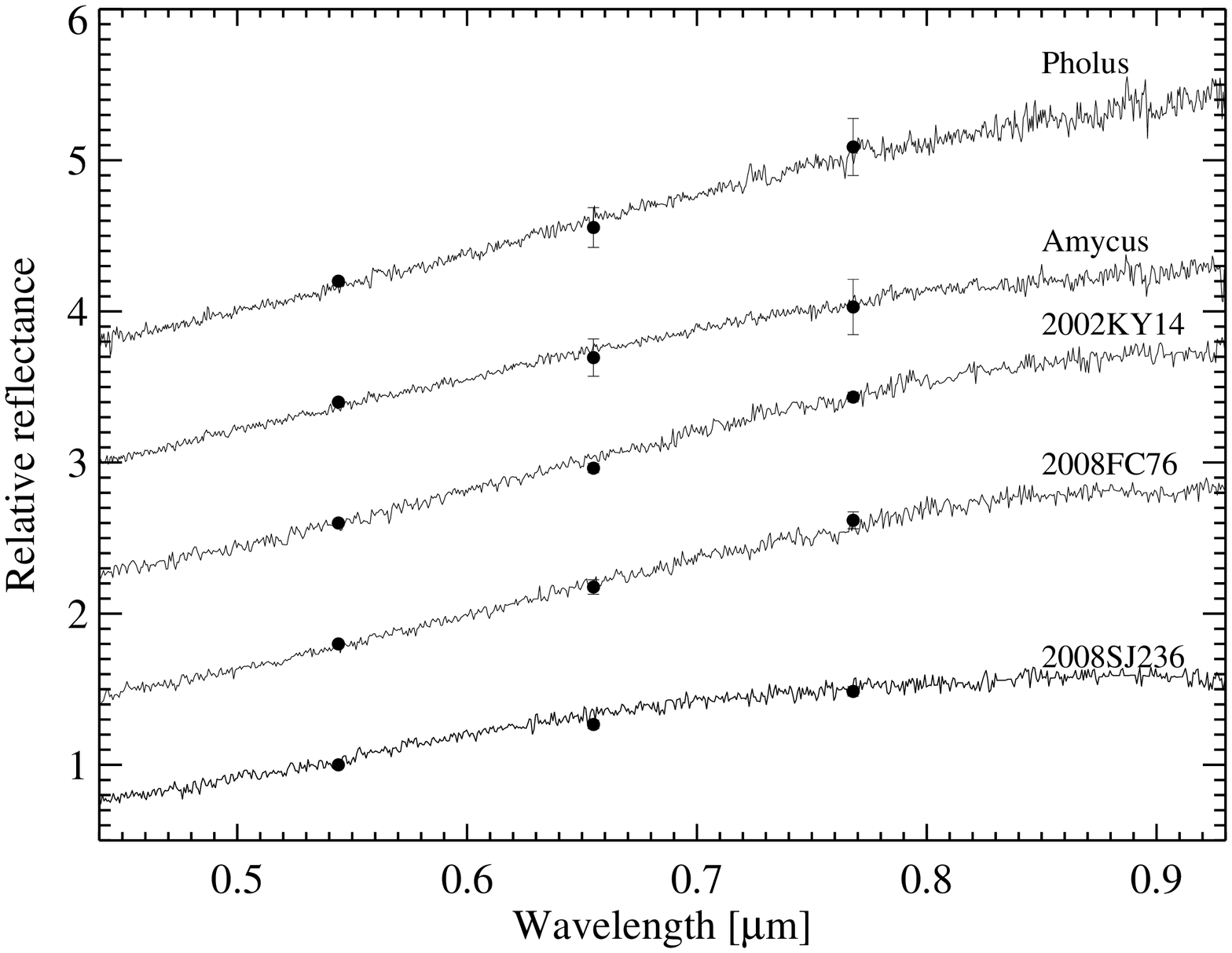}
      \caption{Visible spectra of Centaurs. Spectra are shifted by 0.8 for clarity. The colour indices converted to spectral reflectance are also shown on each spectrum.}
         \label{f2}
   \end{figure}
   
   \begin{figure}
   \centering
\includegraphics[width=7cm,angle=90]{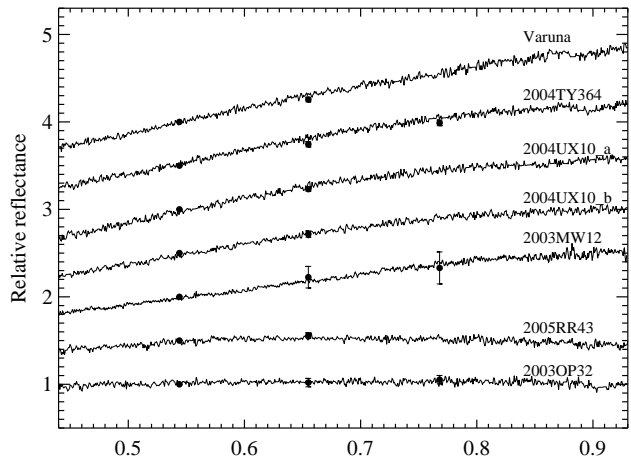}
      \caption{Visible spectra of Classical TNOs. Spectra are shifted by 0.5 for clarity. The colour indices converted to spectral reflectance are also shown on each spectrum. 2004 UX10 was observed during two different nights: the spectrum labelled \textquotedblleft a\textquotedblright~\ was taken on 4 Dec. 2007, the one labelled \textquotedblleft b\textquotedblright~\ on 5 Dec. 2007. }
         \label{f3}
   \end{figure}
   
   \begin{figure}
   \centering
\includegraphics[width=7cm,angle=90]{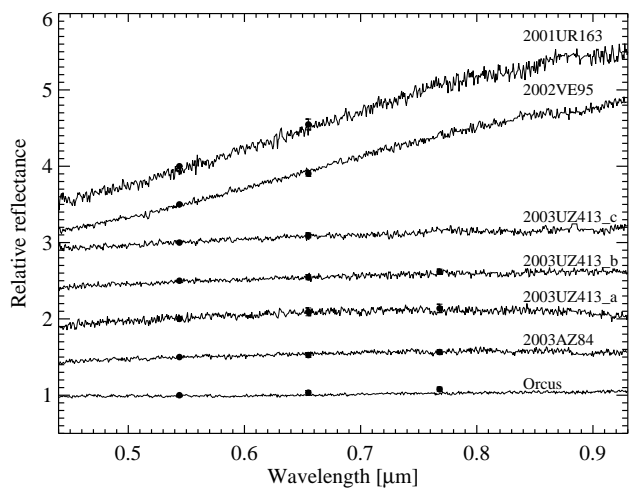}
      \caption{Visible spectra of resonant TNOs. All are plutinos except 42301 (2001 UR163) which is in the 9:4 resonance with Neptune. We observed 2003 UZ413 during 3 different runs: the spectrum labelled \textquotedblleft a\textquotedblright~\ is from Sept. 2008,  \textquotedblleft b\textquotedblright~\ from Nov. 2008, and \textquotedblleft c\textquotedblright~\ from Dec. 2007.  Spectra are shifted by 0.5 for clarity. The colour indices converted to spectral reflectance are also shown on each spectrum.}
         \label{f4}
   \end{figure}
   
     \begin{figure}
   \centering
\includegraphics[width=7cm,angle=90]{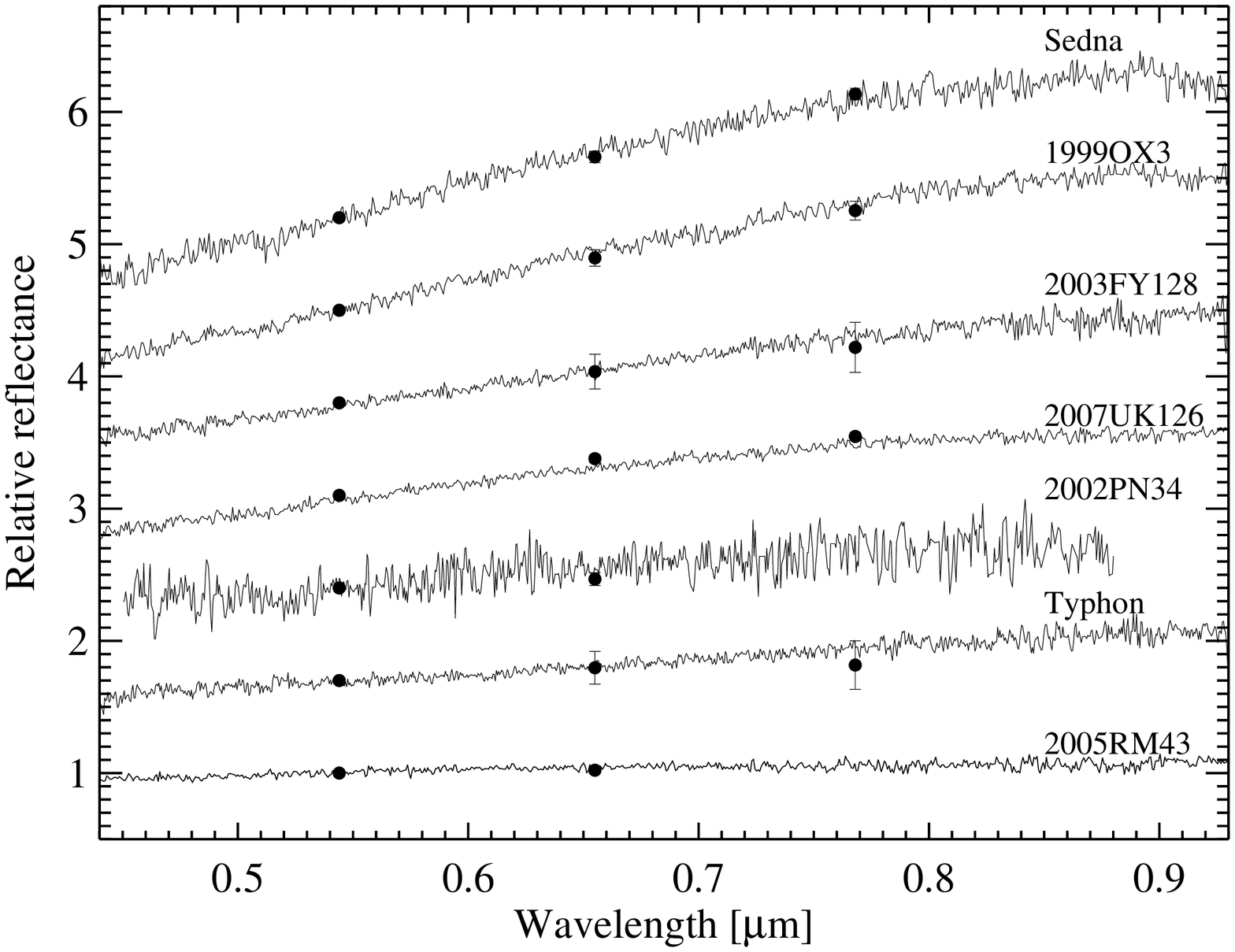}
      \caption{Visible spectra of SDOs and detached objects (Sedna and 120132 2003 FY128). Spectra are shifted by 0.7 for clarity. The colour indices converted to spectral reflectance are also shown on each spectrum.}
         \label{f5}
   \end{figure}
   
We have obtained new spectra of 28 objects, 15 of which were observed in visible spectroscopy for the first time:
10 spectra of Centaurs (Figs.~\ref{f1},~\ref{f2}), 6 of classical TNOs (Fig.~\ref{f3}), 5 of resonants (4 plutinos plus 2001 UR163 which is in the 9:4 resonance with Neptune, Fig.~\ref{f4}), 5 scattered disk objects (SDOs) and two extended SDOs (Fig.~\ref{f5}). We distinguish the dynamical classes according to the Gladman et al. (2008) classification scheme. First of all, the wide variety in the spectral behaviour of both the Centaur and TNO populations is confirmed, since the spectral slopes span a wide range of colours, from grey to very red. All the objects have featureless spectra, with the exception of 10199 Chariklo and 42355 Typhon, whose spectra, showing faint absorption features, have already been investigated by Guilbert et al. (2009) and Alvarez-Candal et al. (2009). They are shown in this paper for completeness.
 
\subsection{Centaurs}
Six out of the 10 Centaurs observed have been spectroscopically observed for the first time (2007 UM126, 2007 VH305, 2008 SJ236, 2008 FC76, 2002 KY14 and 2003 CO1). All spectra are featureless except for 10199 Chariklo (discussed in Guilbert et al. 2009), and their spectral slopes range from 9 to 48 \%/($10^3$\AA) (Figs.~\ref{f1},~\ref{f2}). For Chariklo, an absorption band centred at 0.65 $\mu$m, 0.3 $\mu$m wide and with a depth of 2\% as compared to the continuum was identified on the February 2008 spectrum by Guilbert et al. (2009) and tentatively attributed to the presence of aqueous altered material on its surface. This detection confirms the possible feature of 1\% depth reported by Alvarez-Candal et al. (2008) on a Chariklo spectrum acquired on March 2007.  \\
The Centaurs' spectra are almost linear, but 4 objects (2008 SJ236, 2008 FC76, 2002 KY14 and 55576 Amycus) show a departure from a linear trend for wavelengths longer than 0.75 $\mu$m, as already noticed for some TNOs and Centaurs by  Alvarez-Candal et al. (2008). 

Comparing the spectral slopes of the objects previously observed in the literature (Table~\ref{allslopes}), we found a slightly higher slope value for 55576 Amycus, 10199 Chariklo, and 52872 Okyrhoe, while for 5145 Pholus we find a value similar to the one published by Binzel (1992).
Our spectral slopes are confirmed by the $V-R$ and $V-I$ colour indices obtained just before the spectroscopic observations (Fig.~\ref{f1}, ~\ref{f2}).  For 52872 Okyrhoe and 55576 Amycus, the difference in the spectral slopes are quite small and probably just related to the use of different solar analogue stars and/or to different observing and set-up conditions. For Chariklo, the different spectral behaviour shown both in the visible and near infrared regions was interpreted as coming from to surface heterogeneities (Guilbert et al. 2009).

\subsection{Classical TNOs}
The spectra of 6 classical TNOs are shown in Fig.~\ref{f3}, and all were observed in visible spectroscopy for the first time except 2005 RR43 and 2003 OP32, previously observed by Alvarez-Candal et al. (2008) and Pinilla-Alonso et al. (2008 and 2007). The spectral slopes range from 1 to 25 \%/($10^3$\AA), and the values obtained in this work for 2005 RR43 and 2003 OP32 are in good agreement with those reported in the literature (Table~\ref{allslopes}).
 All the spectra are linear and featureless, and only 2004 UX10 shows a departure from linearity after 0.8 $\mu$m similar to what is seen on some Centaurs. 2004 UX10 was observed twice during the December 2007 run, and the two spectra are identical.     
 
Varuna was observed spectrophotometrically by Lederer and Vilas (1993) with 5 filters in the visible region. The spectral slope value estimated from their Figure 1 is $\sim$ 46 \%/($10^3$\AA), very different from the one obtained here (24.9\%/($10^3$\AA), see Table~\ref{allslopes}). Jewitt \& Sheppard (2002) measured a rotational period of 6.34 hours and significant photometric variation (0.42 mag in R) for Varuna suggesting a triaxial shape (Lacerda \& Jewitt 2007). The different spectral slope values may stem from heterogeneities on the surface of this large TNO.

\subsection{Resonant TNOs}
\begin{figure}
  \centering
\includegraphics[width=6.5cm,angle=90]{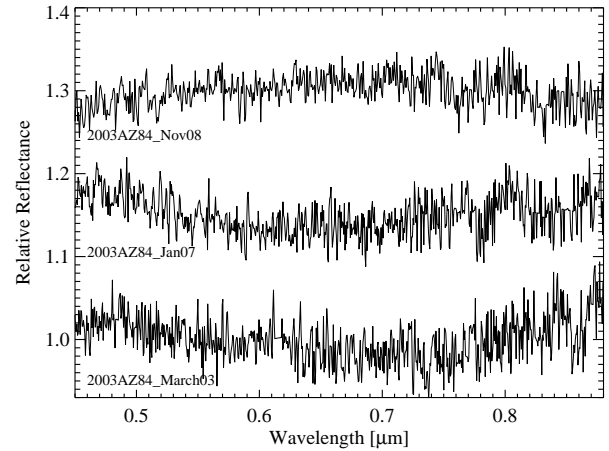}
      \caption{Visible spectra of 2003 AZ84 taken during 3 different observing runs, after continuum slope removal. No indication of the band seen in the March 2003 and Jan. 2007 data appears in the spectrum obtained on Nov. 2008.}
         \label{AZ84}
   \end{figure}

Figure~\ref{f4} shows the spectra of the 5 resonant TNOs observed. All have featureless spectra, and their spectral slope values range from 1 to 40 \%/($10^3$\AA) for the plutinos, while 42301 2001 UR163, populating the 9:4 resonance with Neptune, has a very steep spectral slope (51 \%/($10^3$\AA)). 
Orcus, 2003 AZ84, and 55638 (2002 VE95) were observed previously.  The new spectral slope values 
obtained for Orcus and 2003 AZ84 are in agreement with those in the literature (see Table~\ref{allslopes}), while the values are different for 55638 (2002 VE95) and it is possible that this object has a heterogeneous surface.
Despite the similarity of 2003 AZ84 spectral slope values obtained in different observing runs, the new spectrum of 2003 AZ84 is featureless, while a weak band centred on 0.7 $\mu$m with a depth of about 3\% with respect to the continuum and a width of more than 0.3 $\mu$m was reported by Fornasier et al. (2004a). This feature was also detected by Alvarez-Candal et al. (2008) on the January 2007 data acquired during this LP. \\
We processed the 3 available spectra of 2003 AZ84 (March 2003, January 2007, and November 2008) by removing a linear continuum, computed by a linear fit in the 0.46 and 0.88 $\mu$m wavelength range. The spectra after the continuum removal are plotted in Fig.~\ref{AZ84}: the new data on 2003 AZ84 do not show any absorption band, except for some small features that are clearly residuals of the background removal (in particular 
the O$_{2}$A band around 0.76 $\mu$m and the water telluric bands around 0.72 $\mu$m and 0.83 $\mu$m). It is possible that 2003 AZ84 has a heterogeneous surface composition. Its rotational period is 6.71 or 6.76 hours for a single-peaked solution or 10.56 hours for a double-peaked solution (Sheppard \& Jewitt 2003, Ortiz et al. 2006). A small satellite 5 magnitudes fainter than the primary is reported by Brown and Suer (2007) with HST observations. The satellite flux is only 1\% of the primary flux, assuming a similar albedo value (0.12, Stansberry et al. 2008), so it is unlikely that it would affect the spectral behaviour of 2003 AZ84. 

We observed 2003 UZ413 during three different runs because it is a peculiar TNO with  estimated high density, a lightcurve amplitude of 0.13$\pm$0.03 magnitudes, and a fast rotational period of 4.13$\pm$0.05 hours (Perna et al. 2009b), presently the second fastest rotator among TNOs after 136108 Haumea, which has a rotational period of 3.9155$\pm$0.0001 hours (Lacerda et al. 2008). The 3 spectra were obtained in December 2007, September 2008, and November 2008. Since the December 2007 observations were taken very close in time (1 day apart) to the determination of 2003 UZ413 rotational period, we can derive their position on the lightcurve precisely, that is at 0.64 of the phase curve (see Fig. 5 in Perna et al. 2009b), on the second peak. Using the current accuracy in the rotational period determination, the 2008 September and November observations are located at 0.49 (just after the first minimum) and 0.78 (just before the second minimum) of the phase curve, respectively (but with some uncertainty as the observations were obtained far away from the time of the rotational period determination). Our observations span less than 1/3 of the rotational period and therefore less than 1/3 of the surface. 
The 2008 September and November spectra have very similar spectral slope values (5.0 and 4.9 \%/($10^3$\AA), respectively), while the 2007 December spectrum has a slightly higher value (6.2 \%/ $10^3$\AA). Although the 3 values are within the uncertainties, the higher spectral gradient corresponding to the spectrum acquired at the peak of the lightcurve could indicate surface inhomogeneity.  It is worth noting that the very fast spinning Haumea shows evidence of surface heterogeneity (Lacerda et al. 2008, 2009). Further observations covering the whole rotational period of 2003 UZ413 are needed to confirm that the spectral variation on its surface is real and that it may display surface variability.
 
\subsection{SDOs and Detached objects}
The spectra of the 7 SDOs and detached objects are shown in Fig.~\ref{f5}. The objects 145451 (2005 RM43), 73480 (2002 PN34), and 2007 UK126 were spectroscopically observed for the first time. The spectral slope values range from 2 to 40 \%/$10^3$\AA. 
For the objects previously observed in the literature, there is good agreement for the spectral slope of Sedna,  while for Typhon, 2003 FY128, and above all 44594 (1999 OX3), the values are different (Table~\ref{allslopes}). Our spectra agree with the colours converted to reflectance (Fig.~\ref{f5}), and likewise for the spectra already published in the literature with their associated broadband colours. It is possible that Typhon and 2003 FY128, which have rotational periods estimated to be longer than 5 and 7 hours respectively (Dotto et al. 2008), and 1999 OX3 (period not yet determined) have inhomogeneous surface compositions.   \\
For 42355 Typhon, a faint absorption feature, similar to that identified on Chariklo, was noticed by Alvarez-Candal et al. (2009). This band is centred on 0.62 $\mu$m with a depth of 3\% as compared to the continuum, and is similar to the subtle broad feature reported previously  by Alvarez-Candal et al. (2008) on the 2007 January spectrum of Typhon. The identified shallow broad feature resembles absorption bands detected on some main belt dark asteroids (Vilas et al. 1994; Fornasier et al. 1999), and is attributed to the presence of minerals on their surface produced by the aqueous alteration of anhydrous silicates. 

\section{Discussion}

TNOs and Centaurs reveal an extraordinary spectral variety, as already noticed by several authors from the broadband photometry carried out on a larger sample of objects (Tegler et al. 2008, Doressoundiram et al. 2008 and reference therein). In the visible wavelength range, most of them have featureless and linear spectra, with spectral gradients from neutral--grey to very red. A few of them reveal some absorption bands: two very large bodies (Eris and Makemake, beside Pluto) show strong features due to methane ice (Brown et al. 2008 and reference therein); Haumea might have a band at 0.5773 $\mu$m, possibly due to O$_{2}$ ice that needs to be confirmed (Tegler et al. 2007); and five bodies seem to show faint absorption bands attributed to aqueous alteration processes.
Two of these objects are Chariklo and Typhon, already discussed in the previous section. The other three bodies are 47932 (2000 GN171) and 38628 Huya (2000 EB173) (Lazzarin et al. 2003), and 208996 (2003 AZ84) (Fornasier et al. 2004a). For 47932 and Huya, the feature has never been confirmed (de Bergh et al. 2004; Fornasier et al. 2004a; Alvarez-Candal et al. 2008), even when observing more than half of the rotational period of 47932, and this was interpreted as the result of surface composition heterogeneities during the TNO's rotation. \\
For 2003 AZ84, the weak feature centred around 0.7 $\mu$m reported by Fornasier et al. (2004a) was also detected by Alvarez-Candal et al. (2008) on data acquired during this LP on January 2007, but is not confirmed in these more recent observations (Figs.~\ref{f4} and ~\ref{AZ84}). Again it is possible that the surface of this plutino is not homogeneous. \\
The features identified on these bodies, most of all on Typhon, Chariklo and 2003 AZ84, are very faint and look similar to those seen on some low albedo main belt asteroids that have been attributed to hydrated silicates. How aqueous alteration processes could have occurred in the outer solar System is not well understood, and it is also possible that some hydrated minerals formed directly in the solar nebula. A complete discussion about the possible effect of the aqueous alteration process on TNOs and Centaurs is reported in de Bergh et al. (2004). 
   \begin{figure}
   \centering
\includegraphics[width=10cm,angle=0]{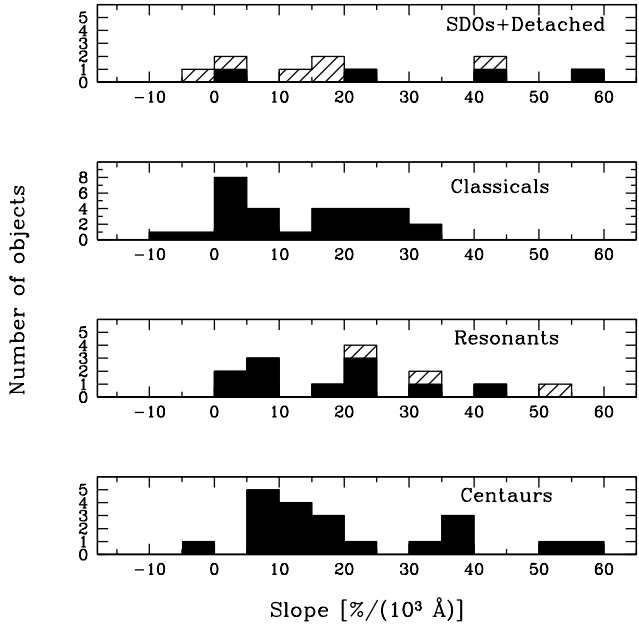}
      \caption{Distribution of TNOs and Centaurs as a function of the
spectral slope. The sample comprises 20 Centaurs, 14 Resonants of which 11 are plutinos (black histogram), and 3 are in the resonances 11:2, 5:2 and 9:4 with Neptune (hatched areas), 29 Classicals, 6 SDOs (hatched areas) and 4 detached bodies.}
         \label{isto}
         \end{figure}

\begin{table}[t]
\caption{Mean spectral slope values and standand deviation of TNOs and Centaurs.}
\label{mean}
\begin{center}
\begin{tabular}{|l|c|c|} \hline
Class & mean & std deviation \\ 
      & [\%/ (10$^3$ \AA)] & [\%/ (10$^3$ \AA)] \\ \hline
Centaurs & 20.5 & 15.1  \\
Classicals & 13.8 & 12.1\\
Resonants  & 20.8 & 15.6\\
SDOs \& Detached  & 21.1 & 18.3  \\
all TNOs \& Centaurs & 17.9 & 14.8 \\ \hline
\end{tabular}
\end{center}
\end{table}

For a better analysis of the spectral slope distribution for Centaurs and TNOs we collected all the visible spectral slopes obtained from spectroscopy available from this work and in the literature (Table~\ref{allslopes}).  Including also the six classical TNOs 86047 (1999 OY3), 86177 (1999 RY215), 181855 (1998 WT31), 1998 HL151, 2000 CG105, and 2002 GH32, whose spectral slope gradient evaluated from spectrophotometry is presented in Ragozzine and Brown (2007, and reference therein), we get a sample of 20 Centaurs and 53 TNOs (14 resonants, 29 classicals, 6 SDO and 4 detached objects, including the dwarf planet Eris). For objects reported in Table~\ref{allslopes} with more than one observation we used a weighted mean of the spectral slope values.  \\
For a clearer analysis of the Centaurs and TNOs spectral slope distribution we show in Fig.~\ref{isto} the number of objects as a function of spectral slope for the 4 distinct dynamical groups: Centaurs, resonants, classicals, and a group including both scattered and detached TNOs. \\
In Fig.~\ref{isto} it is evident that there is a lack of very red objects in the classical TNOs investigated, as all have spectral slopes lower than 35 \%/10$^3$ \AA, and they also show a lower spectral slope mean value compared to the other classes (Table~\ref{mean}). More than half of the classical population (51.7\%) has $S < 13$ \%/10$^3$ \AA\ and is peaked at neutral--grey colours (mean slope value of 3.4 \%/10$^3$ \AA), and the remaining part shows medium-red spectra with a mean slope value of 24.9 \%/10$^3$ \AA. \\  
The Centaurs, resonants, and SDOs--detached objects have very similar mean spectral slope values (Table~\ref{mean}), and most of them have neutral to moderately red slopes ($\sim$ 70\% of the population of each class has $S <$ 24.0 \%/10$^3$ \AA). Our limited dataset based on spectral slope does not clearly show a bimodal distribution for the Centaurs as seen with photometric colours on a larger sample by several authors (Tegler et al. 2008, and references therein). \\ 
The whole sample (TNOs+Centaurs) has a mean slope value of 17.9 \%/10$^3$ \AA. The majority  of the bodies (54 out of 73, that is 74\% of the sample) has $ S < 25$ \%/10$^3$ \AA, 17.8\% of the sample has  $ 25 <S < 40$ \%/10$^3$ \AA, and only 8.2\% has a very red spectral slope value ($ S > 40 $ \%/10$^3$ \AA). About 20\% of the bodies show neutral to grey spectral behaviour ($ -2 < S < 5$ \%/10$^3$ \AA).

It has been claimed that the Kuiper Belt is a possible source of Jupiter Trojans, which would have been formed there and then trapped by Jupiter in the L4 and L5 Lagrangian points during planetary migration (Morbidelli et al. 2005). Fornasier et al. (2007) analysed a sample of 146 Jupiter Trojans (68~L5 and 78~L4) for which visible spectroscopy was available, and calculated an average slope of 9.15$\pm$4.19\%/$10^{3}$\AA\ for objects populating the L5 swarm and 6.10$\pm$4.48\%/$10^3$\AA\ for the L4 ones. A comparison between the spectral slope distribution of Jupiter Trojans (Fig. 14 in Fornasier et al. 2007) and of Centaurs--TNOs (Fig.~\ref{isto}) shows that the Trojans' distribution is very narrow and distinguishable from that of Centaurs-TNOs. The Trojans' spectral gradient is similar only to the neutral-grey to moderately red objects in the Centaurs and TNOs population. The lack of red objects in the Trojan population compared to Centaurs and TNOs might reflect either an intrinsic different planetesimal composition with increasing heliocentric distances or a diverse degree of surface alteration and/or a different collisional history.

\subsection{Statistical analysis}

\begin{table}[t]
\caption{Spearman correlation}
\label{stat}
\begin{center}
\begin{tabular}{|l|c|c|c|} \hline
 Class & $\rho$ & $P_{r}$ & $n$\\ \hline
 {\bf Centaurs} & & & \\
$S$ vs $i$ &   -0.00451293  &   0.98495& 20 \\
$S$ vs $e$ &  0.430989  &  0.0578041& 20\\
$S$ vs $a$ & 0.0691989  &   0.771903& 20\\
$S$ vs $H$ &  0.398494  &  0.0818076& 20\\
$S$ vs $q$ &  -0.0910117  &   0.702757& 20\\
 {\bf Classicals} & & &\\
{\boldmath{$S$ vs $i$}} & {\bf -0.641537}  & {\bf 0.000176407} & {\bf 29} \\
{\boldmath{$S$ vs $e$}} & {\bf -0.510963}  & {\bf 0.00461781}   & {\bf 29} \\
$S$ vs $a$ &  -0.109633 &   0.571303 & 29 \\
$S$ vs $H$ & -0.0892836 &  0.64510& 29 \\
$S$ vs $q$ &  0.430648   & 0.0196961 & 29 \\
 {\bf Resonants} & & &\\
$S$ vs $i$ & -0.393407   &  0.164032 & 14 \\
{\boldmath{$S$ vs $e$}}&  {\bf 0.767033}   & {\bf 0.00136768} & {\bf 14} \\
$S$ vs $a$ &   0.560440   &  0.0371045& 14 \\
$S$ vs $H$ & 0.382838   & 0.176678 & 14 \\
$S$ vs $q$ &   0.142857  &  0.626118 & 14 \\
 {\bf SDOs} & &  &\\
$S$ vs $i$ &   -0.542857  &   0.265703 & 6 \\
{\boldmath{$S$ vs $e$}} &  {\bf -0.885714}   & {\bf 0.0188455} & {\bf 6} \\
$S$ vs $a$ &  -0.657143   &  0.156175 & 6 \\
$S$ vs $H$ & 0.0857143    & 0.871743 & 6 \\
$S$ vs $q$ &  -0.0857143  &   0.871743 & 6 \\
 {\bf Detached Objs} & & &\\
$S$ vs $i$ &   -0.200000  &   0.800000 & 4 \\
$S$ vs $e$ &   0.00000    &  1.00000 & 4 \\
$S$ vs $a$ &  -0.400000   &  0.600000 & 4\\
$S$ vs $H$ & 0.800000     & 0.200000 & 4 \\
$S$ vs $q$ &  -0.400000   &  0.600000 &  4 \\ \hline
\end{tabular}
\end{center}
\begin{list}{}{}
\item Spearman correlation results of the spectral slope $S$, orbital parameters (inclination $i$, eccentricity $e$, semimajor axis $a$, perihelion $q$), and absolute magnitude $H$. $\rho$ is the Spearman rank correlation, $P_{r}$ is the significance level, and $n$ the number of objectsused in the statistical analysis. Strongest correlations are in bold. 
\end{list}
\end{table}

We ran a Spearman rank correlation (Spearman 1904) to look for possible correlations between spectral slope values ($S$) and orbital elements (inclination $i$, eccentricity $e$, and semimajor axis $a$) or absolute magnitude $H$ inside the classical, resonant, and Centaur classes. The function calculating the Spearman correlation gives a two-element vector containing the rank correlation coefficient ($\rho$) and the  two-sided significance of its deviation from zero ($P_{r}$).
The value of $\rho$ varies between -1 and 1: if it is close to zero this means no correlation, if $|\rho|$ is close to 1 then a correlation exists. The significance is a value in the interval $0 < P_{r} < 1$, and a low value indicates a significant correlation.
We consider a strong correlation to have $P_{r} < 0.01 $ and $|\rho| > 0.6 $, and a weak correlation to have $P_{r} < 0.05 $ and $ 0.3 < |\rho| < 0.6 $.

\begin{figure*}[t]
   \centering
\includegraphics[width=12cm,angle=90]{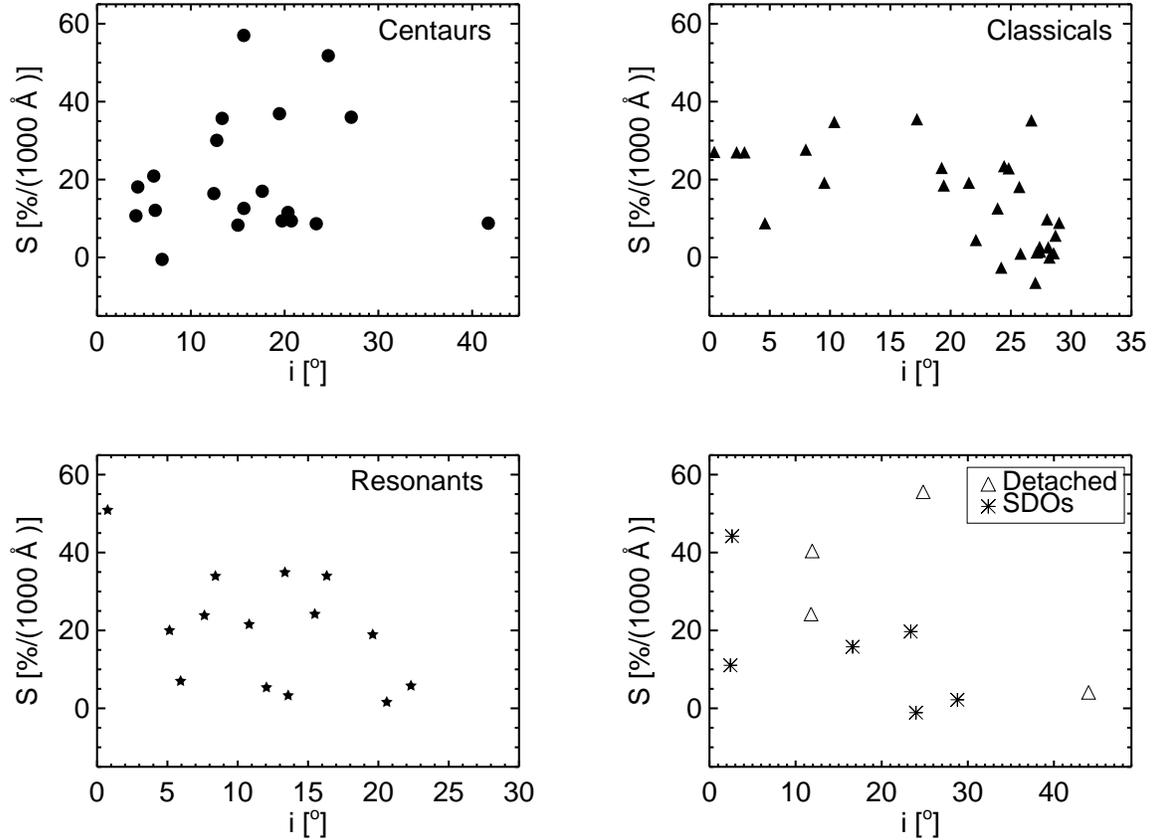}
      \caption{Spectral slope versus orbital inclination. The sample comprises 20 Centaurs, 14 Resonants, 29 Classicals, 6 SDOs, and 4 detached bodies.}
         \label{ff3}
   \end{figure*}

\begin{figure*}[t]
   \centering
\includegraphics[width=12cm,angle=90]{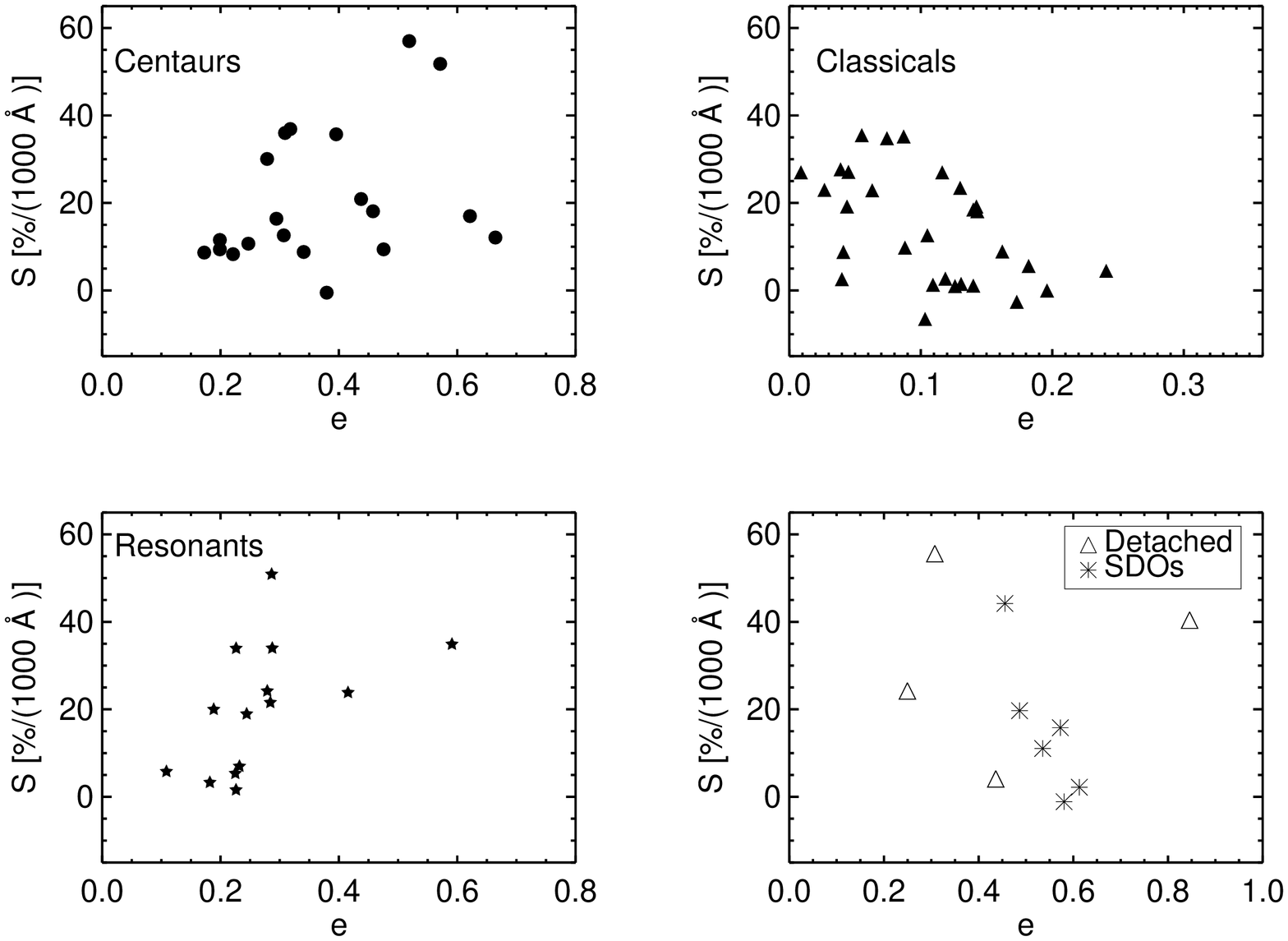}
      \caption{Spectral slope versus orbital eccentricity. The sample comprises 20 Centaurs, 14 Resonants, 29 Classicals, 6 SDOs, and 4 detached bodies.}
         \label{ff2}
   \end{figure*}

For the 29 classical TNOs, we find an anticorrelation between $S$ and $i$ and a weaker anticorrelation between $S$ and $e$ (Table~\ref{stat} and Figs.~\ref{ff3},~\ref{ff2}). The anticorrelation between $S$ and $i$ was mentioned in Tegler \& Romanishin (2000), first investigated by Trujillo and Brown (2002) and then by several authors (see Doressoundiram et al. 2008, and references therein). This strong anticorrelation has also been recently confirmed by Santos-Sanz et al. (2009) and Peixinho et al. (2008) on a sample of 73 and 69 classical TNOs, respectively, investigated in photometry. Nevertheless, Peixinho et al. (2008) find that the optical colours are independent of inclination below $i\sim12^{o}$ and that they provide no evidence of a change in the $B-R$ colour distribution at the boundary between dynamically hot and cold populations (at $i\sim5^{o}$). Our smaller sample of classical TNOs has a gap in the inclination distribution for $\sim 10 < i < 17 ^{o}$. If we consider the 7 objects with $i< 10.4^{o}$, there is effectively no correlation between $S$ and $i$ ($\rho = 0.198206$ and  $P_{r}=0.993299$), while the bodies with $i> 10.4^{o}$ show an anticorrelation between $S$ and $i$ that is weaker than the one found for the entire sample ($\rho = -0.503106$ and  $P_{r}=0.017000$), so our spectral data independently confirm the result by Peixinho et al. (2008) that classical TNOs with $i < 12 ^{o}$ show no correlation between surface colour and inclination. \\
Classical TNOs with $e > 0.14 $ have neutral to grey slopes, and there is a lack of red bodies (Fig.~\ref{ff2}). The anticorrelation between $S$ and $e$ in the classical TNOs might be related to the weak correlation found between $S$ and the perihelion distance $q$ (Table~\ref{stat}). These correlations are mainly driven by the hot population (25 out of the 29 objects in the sample have $i > 5^{o}$), so we can conclude that hot classicals show a neutral/blue spectral behaviour for increasing values of $i$ and $e$, and they are also located at short perihelion distances. Similar results have been obtained by Santos-Sanz et al. (2009) analysing colours versus orbital elements. They interpreted the correlation between perihelion distances and colours as support for colouring scenarios: blue objects may be the results of resurfacing by fresh ices due to sublimation of volatiles or by micrometeoroid bombardment, as these mechanisms are more efficient at smaller perihelion distances. 

The 5 Centaurs in the sample with inclination lower than 10$^{o}$ have $S < 21$ \%/10$^3$ \AA, while for those with higher inclination we distinguish 2 groups: one (9 bodies) with $8 < S < 17$ \%/10$^3$ \AA\, and one (6 bodies) with  $30 < S < 57$ \%/10$^3$ \AA\ (Fig.~\ref{ff3}). Nevertheless, no correlation is found between $S$ and $i$, or with other parameters, except a weak one between $S$ and $e$ (Table~\ref{stat}).
For the resonant TNOs, there is a strong correlation between $S$ and $e$ (Table~\ref{stat}). Objects with high eccentricity tend to be redder than those in low eccentricity orbits; that is, the spectral slope tends to increase with eccentricity. If we exclude the 3 non plutino objects, the correlation becomes weaker ($\rho = 0.681818$ and $P_{r} = 0.020843$). 

The stronger correlation found inside the whole sample is for the SDOs: spectral slope decreases with increasing eccentricity (Fig.~\ref{ff2}), as shown by the anticorrelation between $S$ and $e$ (Table~\ref{stat}), but we must be careful as our sample of 6 SDOs is too small to be statistically significant. This strong correlation was also found by Santos-Sanz et al. (2009) on a larger sample (25) of SDOs. 

Both Centaurs and Resonants tend to have higher slope values with increasing eccentricity, and this trend is opposite to what is seen for SDOs. Analysing the colour-colour correlations, Peixinho et al. (2004) found some resemblance between the Centaur and Plutino distributions, and they also found that the correlations for these two classes is opposite to those of SDOs. They suggest that Centaurs might have mainly originated from the Plutino population, not from SDOs, and that the mechanism changing the surface spectral behaviour might be similar for Centaurs and Plutinos. This interpretation is strengthened by the similar $S - e$  correlation between the Centaurs and Plutinos, and by the $S - e$ anticorrelation for SDOs found here.  
 
\section{Conclusions}
In this work we have presented new visible spectra for 18 TNOs and 10 Centaurs obtained in the framework of a LP devoted to the investigation of these pristine bodies. The main results obtained may be summarised as follows.
\begin{itemize}
\item Our data confirm the diversity in the surface composition of the Centaur and TNO populations, with a spectral slope gradient ranging from $\sim$1 to 51 \%/10$^3$ \AA. 
\item Absorption features: all the new spectra are featureless, excluding 10199 Chariklo and 42355 Typhon, which show faint broadband absorptions centred at 0.62-0.65 $\mu$m, tentatively attributed to phyllosilicates. These two objects are fully discussed in two separate papers (Guilbert et al. 2009, Alvarez-Candal et al. 2009)   
\item The new spectrum obtained of the plutino 2003 AZ84 is featureless, while a faint broadband feature possibly attributed to hydrated silicates was previously reported by Fornasier et al. (2004) and Alvarez-Candal (2008), so we cannot confirm an aqueous alteration action on its surface, but we cannot exclude that 2003 AZ84 simply has a heterogeneous surface composition. 
\item Spectral heterogeneity: comparing the spectral slope values of the objects obtained here with those available in the literature, we found important differences for a few objects, in particular for 10199 Chariklo, 20000 Varuna, 44594 (1999 OX3), and 120132 (2003 FY128). This variability on the spectral behaviour may stem from surface heterogeneities. We observed the fast-rotating (period of 4.13$\pm$0.05 hours, Perna et al. 2009b) plutino 2003 UZ413 during 3 different runs. We found that the slope of the spectrum acquired on the second maximum of the lightcurve is higher than the two slopes obtained near the minima, even if the differences are not greater than the uncertainties. Further observations covering the whole rotational period of 2003 UZ413 are needed to confirm if the spectral variation on its surface is real and if this object may display surface variability.
\item  Spectral slope distribution: we combined the new data presented here with the visible spectra published in the literature (coming mainly from the 2 LPs devoted to TNOs and Centaurs carried out at ESO). The total sample contains 20 Centaurs and 53 TNOs. After analysing the spectral slope distribution of Centaurs, classicals, resonants, and SDOs--detached bodies, we found a lack of very red objects in the classical population, as all have spectral slope values $<$ 35 \%/10$^3$ \AA, with the lowest mean value ($\sim$ 14 \%/10$^3$ \AA) in the dynamical classes investigated. 
The Centaur population is dominated (70\%) by bluish to grey or moderately red objects but also comprises very red objects, with spectral slopes greater than 40\%/10$^3$ \AA. Nethertheless, very red objects are a minority in the Centaur and TNO populations (8\%), while most of them (74\%) have spectral slope values $<$ 25\%/10$^3$ \AA.
\item Statistical analysis: running a Spearman rank order analysis, we confirm a strong anticorrelation between the spectral slope and inclination for the classical TNOs, of a weak anticorrelation between $S$ and $e$, and of a weak correlation between $S$ and the perihelion distance. Our sample mainly contains objects in the hot population ($i> 5^{o}$), which thus show neutral/blue spectral behaviour for increasing values of $i$ and $e$. Nevertheless, we do not observe a change in the slope distribution between the hot and cold populations, but we do find that objects with $ i < 12^{o}$ show no correlation between spectral slope and inclination. This result independently confirms the finding by Peixinho et al. (2008) on the colour--inclination relation for classical TNOs.
We found a strong correlation between $S$ and $e$ for resonant bodies, a similar but weaker one for Centaurs, while for SDOs we confirm the anticorrelation between $S$ and $e$ found by Santos-Sanz et al. (2009), even though it must be noted that our sample is very limited. 
\end{itemize}

\begin{acknowledgements}
We thank an anonymous referee for his/her useful comments in the reviewing process.
\end{acknowledgements}

\setcounter{table}{1}
\begin{center}
\begin{longtable}{|l|c|l|l|}
\caption[]{Spectral slopes values of TNOs and Centaurs.}
\label{allslopes} \\
\hline \multicolumn{1}{|c|}{\textbf{Object}} & \multicolumn{1}{c|}{\textbf{Slope [\%/(\boldmath{$10^{3}$ \AA)}]}} & \multicolumn{1}{c|}{\textbf{Date}} & \multicolumn{1}{|l|}{\textbf{Reference}}  \\ [0.5ex] \hline
   \\[-1.8ex]
\endfirsthead
\multicolumn{4}{c}%
{{\bfseries \tablename\ \thetable{} -- continued from previous page}} \\ \hline
\multicolumn{1}{|c|}{\textbf{Object}} & \multicolumn{1}{c|}{\textbf{Slope [\%/(\boldmath{$10^{3}$ \AA)}]}} & \multicolumn{1}{c|}{\textbf{Date}} & \multicolumn{1}{|c|}{\textbf{Reference}}  \\ \hline 
\endhead
\hline \multicolumn{4}{r}{{Continued on next page}} \\ 
\endfoot
\hline \hline
\endlastfoot
{\bf Centaurs} & & & \\ 
2060 Chiron       &-0.5$\pm$0.2& 1998 Mar. 29  & Barucci et al. 1999     \\
2060 Chiron       &-0.5$\pm$0.2& 1991 Mar. 07  & Fitzsimmons et al. 1994 \\
5145 Pholus       &55.0$\pm$2.0& 1992 Mar. 12  & Binzel 1992              \\
5145 Pholus       &48.6$\pm$0.7& 2008 Apr.12   & This paper               \\ 
7066 Nessus       &57.9$\pm$0.9& 1998 Mar. 30  & Barucci et al. 1999      \\
8405 Asbolus      &17.0$\pm$0.4& 1998 Mar. 30  & Barucci et al. 1999      \\
10199 Chariklo    & 7.8$\pm$0.1& 1998 Mar. 29  & Barucci et al. 1999      \\
10199 Chariklo    & 7.8$\pm$0.5& 2007 Mar. 20  & Alvarez et al. 2008      \\
10199 Chariklo    &10.4$\pm$0.6& 2008 Feb. 04  & This paper               \\
10370 Hylonome    &10.8$\pm$1.1& 1998 Mar. 29  & Barucci et al. 1999      \\
32532 Thereus     &11.1$\pm$0.2& 2001 Oct. 10  & Lazzarin et al. 2003    \\
32532 Thereus     &12.0$\pm$0.7& 2007 Sep. 19  & Alvarez et al. 2008      \\
54598 Bienor      &10.1$\pm$0.2& 2001 Oct. 15  & Lazzarin et al. 2003    \\    
54598 Bienor      & 8.7$\pm$0.7& 2007 Sep. 18  & Alvarez et al. 2008      \\
52872 Okyrhoe     &11.8$\pm$0.2& 2001 Oct. 15  & Lazzarin et al. 2003    \\   
52872 Okyrhoe     &13.4$\pm$0.6& 2008 Feb. 04  & This paper               \\ 
55576 Amycus      &34.3$\pm$0.1& 2003 Mar. 09  & Fornasier et al. 2004a   \\ 
55576 Amycus      &37.1$\pm$0.9& 2008 Apr. 12  & This paper              \\
60558 Echeclus    &23.7$\pm$0.6& 2001 Apr. 27  & Lazzarin et al. 2003    \\  
60558 Echeclus    &12.5$\pm$0.7& 2007 May  14  & Alvarez et al. 2008      \\
63252 2001BL41    &16.4$\pm$0.2& 2002 Feb. 17  & Lazzarin et al. 2003    \\  
83982 Crantor     &35.8$\pm$0.1& 2003 Mar. 09  & Fornasier et al. 2004a   \\
83982 Crantor     &38.5$\pm$1.2& 2007 Jul. 14  &  Alvarez et al. 2008      \\
83982 Crantor     &39.9$\pm$1.3& 2007 Jul. 16  & Alvarez et al. 2008      \\
95626 (2002 GZ32) & 8.3$\pm$0.1& 2003 Mar. 09  & Fornasier et al. 2004a   \\    
120061 (2003 CO1) & 9.4$\pm$0.6& 2008 Apr. 12  & This paper              \\
2002 KY14         &36.9$\pm$0.7& 2008 Sep. 21  & This paper             \\
2007 UM126        & 8.8$\pm$0.7& 2008 Sep. 21  & This paper             \\
2007 VH305        &12.1$\pm$0.7& 2008 Nov. 22  & This paper             \\
2008 FC76         &36.0$\pm$0.7& 2008 Sep. 21  & This paper             \\
2008 SJ236        &20.9$\pm$0.8& 2008 Nov. 22  & This paper             \\
{\bf Classicals} & & & \\ 
19308 (1996 TO66) &  3.0$\pm$3.0 & 1999 Dec. 03   & Boehnhardt et al. 2001 \\
20000 Varuna      & 24.9$\pm$0.6 & 2007 Dic. 04	  & This paper \\
20000 Varuna      & 46.0$\pm$1.0 & 2002 Mar. 30	  & Lederer and Vilas 2003 \\
24835 (1995 SM55)    &-12.0$\pm$3.0 & 1999 Dec. 02 & Boehnhardt et al. 2001 \\
24835 (1995 SM55)     & -0.9$\pm$2.0 & 2007 Sep. 18  & Pinilla-Alonso et al. 2008 \\
35671 (1998 SN165)   &  8.8$\pm$0.4 & 2002 Aug. 12  & Fornasier et al. 2004a   \\  
50000 Quaoar      & 26.9$\pm$0.2 & 2003 Mar 09	  & Fornasier et al. 2004a   \\
50000 Quaoar      & 28.5$\pm$0.8 & 2007 Jul. 15   & Alvarez et al. 2008      \\
55565 (2002 AW197)   & 22.1$\pm$0.2 & 2003 Mar 09	  & Fornasier et al. 2004a   \\
55565 (2002 AW197)   & 24.8$\pm$0.7 & 2007 Jan. 23	  & Alvarez et al. 2008      \\
55636 (2002 TX300)   &  1.0$\pm$0.0 &  2003 Sep. 23 	  & Licandro et al. 2006a \\
55637 (2002 UX25)    & 18.5$\pm$1.1 & 2007 Sep. 18	  & Alvarez et al. 2008      \\
58534 Logos       & 27.0$\pm$5.0 & 1999 May 15	  &  Boehnhardt et al. 2001 \\
79360 (1997 CS29)    & 27.0$\pm$3.0 & 1999 Dec. 05	  &  Boehnhardt et al. 2001 \\
119951 (2002 KX14   & 27.1$\pm$1.0 & 2007 Jul. 13	  &  Alvarez et al. 2008      \\
120178 (2003 OP32)   &  0.9$\pm$0.7 & 2008 Sep. 21	  & This paper \\
120178 (2003 OP32)   &  1.7$\pm$2.0 & 2007 Sep. 15	  & Pinilla-Alonso et al. 2008 \\
120347 (2004 SB60)   & 12.6$\pm$2.0 & 2007 Sep. 16  & Pinilla-Alonso et al. 2008 \\
120348 (2004 TY364)   & 22.9$\pm$0.7 & 2008 Nov. 22   & This paper \\
136108 Haumea     &  0.0$\pm$2.0 & 2005 Aug. 01	  & Pinilla-Alonso et al. 2009 \\
136472 Makemake   &  8.9$\pm$1.0 & 2005 Aug. 01	  & Licandro et al. 2006b \\
144897 (2004 UX10)   & 20.7$\pm$0.8 & 2007 Dec. 04	  & This paper      \\
144897 (2004 UX10)   & 19.2$\pm$0.9 & 2007 Dec. 05	  & This paper      \\
145452 (2005 RN43)   & 23.0$\pm$1.1 & 2007 Sep. 19	  & Alvarez et al. 2008      \\
145453 (2005 RR43)   &  0.6$\pm$0.7 & 2007 Sep. 18	  & Alvarez et al. 2008      \\
145453 (2005 RR43)   &  1.6$\pm$0.6 & 2007 Dec. 04   &     This paper  \\
145453 (2005 RR43)   & -0.4$\pm$2.0 & 2006 Sep. 23	  & Pinilla-Alonso et al. 2007 \\
174567 (2003 MW12)   & 19.2$\pm$0.6 & 2008 Apr. 12	  & This paper      \\
202421 (2005 UQ513)   & 18.1$\pm$2.0 &  2007 Sep. 16	  & Pinilla-Alonso et al. 2008 \\
2003 QW90          & 34.8$\pm$1.0 & 2006 Nov. 09	  &  Alvarez et al. 2008      \\
2003 UZ117         &  1.5$\pm$0.7 & 2006 Dec. 25   & Alvarez et al. 2008      \\
2003 UZ117         & -2.1$\pm$2.0 &  2007 Sep. 17	  &Pinilla-Alonso et al. 2008 \\ 
{\bf Resonants} & & & \\ 
15789 1993SC   & 20.0$\pm$9.0& 1994 Sep. 29 & Luu \& Jewitt 1996 \\
26181 (1996 GQ21) & 34.9$\pm$0.3& 2002 Feb. 17 & Lazzarin et al. 2003 \\
26375 (1999 DE9)  & 25.5$\pm$0.2& 2002 Feb. 17 & Lazzarin et al. 2003 \\
26375 (1999 DE9)  & 22.2$\pm$0.6& 2007 Jan. 22 & Alvarez et al. 2008      \\
28978 Ixion    & 17.9$\pm$0.4& 2003 May 05 & Marchi et al. 2003 \\
28978 Ixion    & 20.0$\pm$0.8& 2007 Jul. 15 &  Alvarez et al. 2008      \\
38628 Huya     & 24.2$\pm$0.4& 2001 Apr. 27 & Lazzarin et al. 2003 \\
42301 (2001 UR163)& 50.9$\pm$0.7& 2007 Dic. 05& This paper      \\
47171 (1999 TC36) & 30.6$\pm$0.2& 2001 Oct 15 & Lazzarin et al. 2003 \\
47171 (1999 TC36) & 37.3$\pm$0.7& 2006 Nov. 09 & Alvarez et al. 2008      \\
47932 (2000 GN171)& 24.1$\pm$0.2& 2003 Apr. 12 & Fornasier et al. 2004a \\
47932 (2000 GN171)& 20.7$\pm$0.2& 2001 Apr. 27 &  Lazzarin et al. 2003 \\
47932 (2000 GN171)& 22.3$\pm$0.7& 2007 Jan. 23 & Alvarez et al. 2008      \\
47932 (2000 GN171)& 21.5$\pm$0.7& 2007 Jan. 24 & Alvarez et al. 2008      \\
47932 (2000 GN171)& 21.8$\pm$0.8& 2007 Mar. 24 & Alvarez et al. 2008      \\
55638 (2002 VE95) & 40.0$\pm$0.7& 2007 Dic. 05 & This paper      \\
55638 (2002 VE95) & 27.9$\pm$0.1& 2004 Oct. 23 & Barucci et al. 2006 \\
90482 Orcus    &  1.6$\pm$0.6& 2008 Feb. 04 & This paper      \\
90482 Orcus    &  1.8$\pm$0.2& 2004 Feb. 29 & Fornasier et al. 2004b \\
91133 (1998 HK151)&  7.0$\pm$3.0& 1999 May 15 & Boehnhardt et al. 2001 \\
2001 QF298      &  5.8$\pm$0.4& 2002 Aug. 12 & Fornasier et al. 2004a \\
2003 AZ84       &  3.3$\pm$0.2& 2003 Mar. 09 &    Fornasier et al. 2004a \\
2003 AZ84       &  3.0$\pm$0.6& 2007 Jan. 24 &  Alvarez et al. 2008      \\
2003 AZ84       &  3.6$\pm$0.6& 2008 Nov. 22 &  This paper      \\
2003 UZ413      &  6.2$\pm$0.6& 2007 Dec. 04 & This paper      \\
2003 UZ413      &  5.0$\pm$0.8& 2008 Sep. 21 & This paper      \\
2003 UZ413      &  4.9$\pm$0.7& 2008 Nov. 22 & This paper      \\
{\bf SDOs} & & & \\
15874 (1996 TL66) & -1.1$\pm$0.7&  1997 Sep. 30   & Luu \& Jewitt 1998 \\
42355 (Typhon)   & 10.0$\pm$0.6&  2007 Jan. 24 & Alvarez et al. 2008      \\
42355 (Typhon)   & 12.1$\pm$0.8&  2008 Apr.  12 & This paper \\
44594 (1999 OX3)  & 52.5$\pm$0.7&  2001 Apr. 27  &  Lazzarin et al. 2003 \\
44594 (1999 OX3)  & 36.2$\pm$0.8&  2008 Sep. 21 &  This paper \\
73480 (2002 PN34) & 15.8$\pm$0.7&  2007 Nov. 10 &   This paper \\
145451 (2005 RM43)&  2.2$\pm$0.7&  2007 Dec. 04 &   This paper \\
2007 UK126      & 19.6$\pm$0.7&  2008 Sep. 21 & This paper \\
{\bf Detached Objects} & & & \\ 
40314 (1999 KR16)  &55.6$\pm$0.7&  2001 Apr. 27  & Lazzarin et al. 2003 \\
90377 Sedna        &40.2$\pm$0.9&  2008 Sep. 21 & This paper      \\
90377 Sedna        &38.3$\pm$1.1&  2004 Oct. 23 & Barucci et al. 2005 \\      
120132 (2003 FY128) &21.7$\pm$0.7&  2007 Jan. 22 &  Alvarez et al. 2008      \\
120132 (2003 FY128) &26.7$\pm$1.0&  2008 Apr. 12 & This paper      \\
136199 Eris      &4.4$\pm$0.7&  2006 Oct. 20 & Alvarez et al. 2008      \\
136199 Eris      &3.8$\pm$0.7&  2007 Sep. 19 & Alvarez et al. 2008      \\ 
\end{longtable}
\end{center}
\begin{list}{}{}
\item Spectral slopes values of all the TNOs and Centaurs observed during this LP at ESO telescopes (presented in this paper and Alvarez et al. 2008). We report also the spectral slope values obtained by spectroscopy available in the literature.
\end{list}

\end{document}